\documentstyle[12pt,aps,floats,prc,epsfig,tighten]{revtex}
\def\Journal#1#2#3#4{{\em #1} {\bf #2}, #3 (#4) }
\def\NPA{{ Nucl. Phys.} A}
\def\PRL{Phys. Rev. Lett.}

\def\PREP{ Phys. Rep.}
\def\PRC{{Phys. Rev.} C}
\def\PL {Phys. Lett.}   
\newcommand{\matrixel}[3]{\mbox{$\left<#1|#2|#3\right>$}}                                                   
\begin{document}
\title{ $\Delta(1232)$ Isobar Excitations and the Ground State of Nuclei}
\author{T. Frick, S. Kaiser,  H. M\"uther}
\address{Institut f\"ur
Theoretische Physik, \\ Universit\"at T\"ubingen, D-72076 T\"ubingen, Germany}
\author{ A. Polls}
\address{Departament d'Estructura i Constituents de la Mat\`eria,\\
Universitat de Barcelona, E-08028 Barcelona, Spain}
\author{D. R. Entem\footnote{On leave from University of
Salamanca, Spain.}, R. Machleidt}
\address{Department of Physics, University of Idaho, Moscow, Idaho 83844,
U.S.A.}
\maketitle
 
\begin{abstract}
The influence of $\Delta$ isobar components on the ground state properties of
nuclear systems is investigated for nuclear matter as well as finite nuclei.
Many-body wave functions, including isobar configurations, and binding energies
are evaluated employing the framework of the coupled-cluster theory. It is
demonstrated that the effect of isobar configurations depends in a rather
sensitive way on the model used for the baryon-baryon interaction. As examples
for realistic baryon-baryon interactions with explicit inclusion of isobar
channels we use the  local ($V28$) and non-local meson exchange potentials
(Bonn$_{2000}$) but also a model recently developed by the Salamanca group,
which is based on a quark picture. The differences obtained for the nuclear
observables are related to the treatment of the interaction, the $\pi$-exchange
contributions in particular, at high momentum transfers.
\end{abstract}
\section{Introduction\label{Introduction}}                                
The development of efficient computing facilities has enabled very sophisticated
calculations for the solution of the nuclear many-body problem. Starting from
realistic models for the nucleon-nucleon (NN) interaction, which give very
accurate fits of the NN scattering data below the threshold for pion production
\cite{cdb,argv18,nijm1},
one can solve the few-nucleon problem up to $A=8$ nucleons in a way which yields
essentially the exact solution\cite{wirin1}. Introducing an additional
three-nucleon force\cite{piep3,urban3} one can obtain results for the basic
low-energy properties of these nuclei, which are in good agreement with the
experimental data. 

This demonstrates that the low-energy properties of nuclei are well described
within the conventional model of nuclear physics, in which nuclei are considered
as a system of nucleons, treated as inert particles interacting via two-body
forces. All sub-nucleonic degrees of freedom, which may lead to
modifications of the hadrons in the nuclear medium, and dynamical
relativistic\cite{serot} effects are represented by a phenomenological
three-nucleon force. 

On the other hand, however, one knows that nucleons cannot really be considered 
as elementary particles and sub-nucleonic degrees of freedom, like e.g.~the
possibility to excite strongly interacting nucleons, could be very important. In
particular the excitation of nucleons to the $\Delta(3,3)$ resonance may have
some effect on the low-energy and bulk properties of nuclear systems. First
investigations on the importance of isobar degrees of freedom have been
performed more than 20 years ago\cite{weber,green,anast,manz}.  Those studies
demonstrated that isobar configurations yield an important contribution to the
medium range attraction of the NN interaction. Conventional models of the NN
interaction account for these mutual polarisation of the interacting nucleons in
a phenomenological way. For example, a part of the $\sigma$ meson exchange in
One-Boson-Exchange (OBE) models for the NN interaction can be related to such
isobar terms\cite{elster}.  

In a conventional nuclear structure calculation this part of the NN interaction
is identical in the nuclear medium as compared to the vacuum where the effective
NN interaction has been adjusted to describe the NN scattering data. If,
however, the isobar degrees of freedom are taken into account explicitely, one
obtains a modification of the $N\Delta$ and $\Delta\Delta$ propagator in the
medium. This implies that the effective NN interaction including such
intermediate isobar states is different in nuclear matter as compared to the
vacuum case. The corresponding part of the medium range attraction is quenched.
This feature has been investigated by various groups using the
Brueckner-Hartree-Fock approximation\cite{green,anast,manz} or within a lowest
order variational calculation\cite{irvine} and binding energies were obtained,
which were much weaker than in corresponding calculations ignoring the explicit
treatment of isobar excitations. 

For nuclear matter at higher densities the explicit consideration of isobar
configurations leads to an enhancement of the pion propagator which has been
called a precursor phenomenon to a phase transition of pion
condensation\cite{pico1,pico2,pico3}. This leads to rather attractive
contributions to the binding energy which originate from ring diagrams involving
$\Delta$-hole excitations\cite{dring}. Nuclear structure studies including
isobar excitations have furthermore been performed for few nucleon
systems\cite{hajduk,picles,schiav}.    

Most of these older studies have been performed using rather simple models for
the baryon-baryon interaction. The transition potentials describing the $NN \to
N\Delta$ and $NN\to \Delta\Delta$ were approximated in terms of local
$\pi$-exchange potentials. During the last years new models for the
baryon-baryon interaction have been developed. It has been demonstrated that a
local approximation of the $\pi$-exchange term tends to overestimate these
contributions considerably\cite{local,mupo}. This is also true for the
transition potentials leading to isobar excitations\cite{frick1}. So it is one
aim of the present investigation to update nuclear structure studies with
explicite treatment of isobar excitations using modern models for the NN
interaction. We are going to compare results for nuclear matter and finite
nuclei, calculated for the Argonne $V28$ potential\cite{V28}, a recent update
of the non-local meson-exchange potential denoted as
Bonn$_{2000}$\cite{mac00} and a model which has recently been developed by
the Salamanca group\cite{entem}. This Salamanca interaction is derived in the
framework of the Chiral Quark Cluster (CQC) model. The problem of two
interacting clusters (baryons) of quarks is solved by means of the resonating
group method. The Pauli principle between the interacting quarks is an
important source for the short-range repulsion of the NN
interaction\cite{lueb}. At large distances the $\pi$ exchange between the
quarks in the two clusters evolves to the $\pi$ exchange between two baryons.
At shorter distances, however, this non-local model for the baryon-baryon
interaction might yield results that are quite different from a meson-exchange
picture. This Salamanca potential does not give such a perfect fit to the NN
scattering phase shifts as the Bonn$_{2000}$ or the $V28$. For the $^1S_0$,
$^1P_0$  and
$^3S_1-^3D_1$ partial waves of the NN system, however, the agreement with the
empirical data is rather good. 

An explicit evaluation of isobar components in the nuclear wave function is also
motivated from recent experiments, which try to measure such isobar
components\cite{exp98}.

The isobar components in the nuclear wave function and the resulting
ground-state properties will be evaluated in an extension of the coupled
cluster method\cite{kuem}. This extension is presented in section 2 where we
will also compare predictions of the coupled-cluster method with the
Brueckner-Hartree-Fock approximation. Results for the binding energy and isobar
probabilities obtained for nuclear matter and finite nuclei will be presented
in section 3. Special attention will be paid to the difference between the
various interaction models. 

\section{Coupled-Cluster Approach with Isobar Excitations}
In the coupled-cluster approach\cite{kuem}  one starts assuming
an appropriate Slater determinant $\Phi$ as a reference state for the system
under consideration. In the examples considered below this reference state
will be a Slater determinant defined in terms of appropriate oscillator
single-particle wave functions for the case of $^{16}$O, while for the case of
infinite nuclear matter $\Phi$ stands for the antisymmetrized wave function
build in terms of plane waves with momenta less than the Fermi momentum $k_F$.
The exact eigenstate $\Psi$ is then written as
\begin{equation}
\Psi = e^S \Phi \label{eq:esansatz}\mbox{,}
\end{equation}
with $S$ being an operator of the form
\begin{equation}
S = \sum_{n=1}^A S_n \label{eq:sumsn}\mbox{,}
\end{equation} 
where $S_n$ is an $n$-particle operator and in order to be complete one has to
consider operators up to $n=A$ with $A$ the number of baryons in the system. The
operator $S_n$ describes the formation of an $n$-particle $n$-hole excitation
relative to the reference state $\Phi$. For the case of $n=2$ it can be written
\begin{equation}
S_2 = \frac{1}{4}\sum_{\nu_1,\nu_2, \rho_1, \rho_2}
\left<\rho_1 \rho_2 \vert S_2 \vert \nu_1\nu_2 \right> a^\dagger_{\rho_1}
a^\dagger_{\rho_2} a_{\nu_2}  a_{\nu_1} \label{eq:sndef}\mbox{.}
\end{equation}
In this equation $a^\dagger_{\rho_i}$ stand for fermion creation operators in
states which are unoccupied in $\Phi$, while  $a_{\nu_i}$ represent
annihilation operators for the nucleon single-particle states which are occupied
in the Slater determinant $\Phi$. Note that the $a^\dagger_{\rho_i}$ may also
represent the creation of $\Delta$ isobar states. Therefore the $S_2$ amplitudes
describe two-particle two-hole excitations relative to $\Phi$ but also $N\Delta$
and $\Delta\Delta$ excitations. 

One can now use the Schr\"odinger equation in the form
\begin{equation}
e^{-S} H e^S \Phi = E \Phi \,,\label{eq:schroed} 
\end{equation}
and project this equation on the reference state $\Phi$ and $n$-particle
$n$-hole states relative to $\Phi$ which we will identify by
$\Phi_{\rho_1...\rho_n\nu_1...\nu_n}$. This leads to an expression for the
energy
\begin{eqnarray}
E & = & \left< \Phi \vert e^{-S} H e^S \vert \Phi \right> \nonumber \\
& = & \left< \Phi \vert H \left(1+S_1 + \frac{1}{2}S_1^2 + S_2 \right) 
\vert \Phi \right>\,, \label{eq:ener}
\end{eqnarray}
and to a set of coupled equations for the amplitudes of linked $n$-particle 
$n$-hole excitations $S_n$. This set of equations is truncated by assuming that 
amplitudes $S_n$ with $n$ larger than a given value $m$ can be ignored. As an
example we consider the $m=2$ approximation, i.e. we ignore the effects of
linked 3-particle 3-hole excitations and higher, and in order to simplify the
notation, we furthermore assume that we have chosen the reference state such that
$S_1$ vanishes (note that this is true in particular for infinite nuclear matter
because of the the translational symmetry). In this case we can write the
correlated two-body state as
$$
\chi_2 \vert \nu_1\nu_2 >_A = (1+S_2) \vert\nu_1\nu_2 >_A\,,
$$
where a subscript $A$ is used to identify antisymmetrized two-body states.
With these simplifications, the
equation for the amplitudes $S_2$ can be reduced to
\begin{eqnarray}
\label{eq:trunc}
\lefteqn{{\matrixel{\rho_1\rho_2}{(T_1+T_2)S_2}{\nu_1\nu_2}}_A-
\sum_{\nu}\Big\{{\matrixel{\rho_1\rho_2}{S_2}{\nu\nu_2}}_A
\matrixel{\nu}{h}{\nu_1}+
{\matrixel{\rho_1\rho_2}{S_2}{\nu_1\nu}}_A
\matrixel{\nu}{h}{\nu_2}\Big\}}
\hspace{2.8cm}
\nonumber
\\
&&+{}{\matrixel{\rho_1\rho_2}{V\chi_2}
{\nu_1\nu_2}}_A+
\underbrace{\frac{1}{2}\sum_{\nu\nu^{\prime}}
{\matrixel{\rho_1\rho_2}{S_2}{\nu\nu^{\prime}}}_A
{\matrixel{\nu\nu^{\prime}}{V\chi_2}{\nu_1\nu_2}}_A}_{(\ast)}=0
\nonumber
\\
\end{eqnarray}      
Note that the term identified with $(\ast)$ includes a summation over intermediate 
hole states. If we ignore this term and use furthermore that the
single-particle hamiltonian of nuclear matter is diagonal in the plane wave
states, $\matrixel{\alpha}{h}{\beta}=\epsilon_{\alpha}\delta_{\alpha\beta},$
we can rewrite (\ref{eq:trunc}) into
\begin{eqnarray}
\label{conn2G}
V\chi_2{\left|\nu_1\nu_2\right>}_A&=
&V{\left|\nu_1\nu_2\right>}_A+
V\frac{Q_{P}}{\underbrace{\epsilon_{\nu_1}+\epsilon_{\nu_2}}_{=\omega}
-T_1-T_2}V\chi_2{\left|\nu_1\nu_2\right>}_A
\nonumber\\
&=&V{\left|\nu_1\nu_2\right>}_A+
VS_2{\left|\nu_1\nu_2\right>}_A
\mbox{.}
\end{eqnarray}
In this equation we have introduced the starting energy $\omega$ and the Pauli
Operator $Q_P$, which restricts the sum over intermediate states to those of the
form $\vert \rho_1\rho_2>$. This means to nucleon single-particle states
$\rho_i$ which are unoccupied in the reference state $\Phi$ or to isobar
excitations. If we identify $V\chi_2$ with the Brueckner $G$-matrix, the 
equation (\ref{conn2G}) takes the form of the Bethe-Goldstone equation
\begin{equation}
G(\omega) = V + V\frac{Q_{P}}{\omega-H_0}G(\omega)\,,\label{eq:betheg}
\end{equation}
with $H_0$ the operator of the kinetic energy, i.e.~assuming the conventional
choice for the spectrum of the particle states\cite{mupo,jeuk,baldo}. 

In order to visualize the relevance of the hole-hole term ($\ast$) in
(\ref{eq:trunc}) we have performed calculations for nuclear matter with and
without this term, using two different models for the NN interaction, which do
not include isobar degrees of freedom, explicitely. One of these examples is the
Argonne V14 potential\cite{V28}, which is defined in terms of 14 operators, each
of them multiplied with a local potential. The second example is the charge
dependent Bonn potential\cite{cdb}, a meson exchange interaction, which is
evaluated in momentum space and contains non-local contributions. 

\begin{figure}
\begin{center}
\epsfig{figure=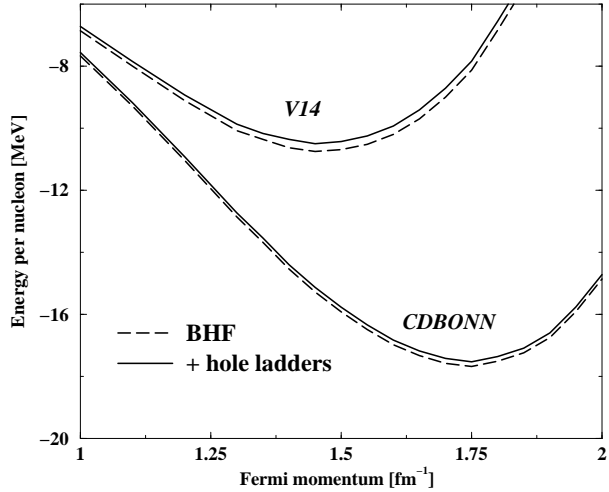,width=8cm}
\end{center}
\caption{Binding energy of nuclear matter as a function of the Fermi momentum.
Results are given for the Argonne V14 and the CDBONN interaction, using the BHF
approximation (dashed lines) and with inclusion of hole-hole ladder terms (solid
lines).\label{fig1}}
\end{figure}

The results of the energy of nuclear matter as a function of the Fermi momentum
$k_F$, calculated in the Brueckner-Hartree-Fock (BHF) approximation is displayed
by the dashed lines in Fig.~\ref{fig1}. The differences originating from the
two interaction models have been discussed e.g. in \cite{local}: A local
interaction (V14) tends to be stiffer than a NN interaction 
based on the non-local meson exchange model (CDBONN), fitting the same NN
scattering data. Since the two-body correlations in nuclear matter are quenched
as compared to the case of NN scattering in the vacuum, a
stiffer interaction tends to predict less binding energy than a softer one. As a
consequence, the BHF energy calculated for the V14 interaction is much less
attractive than for CDBONN. 

If also one includes the hole-hole ladder terms in solving eq.(\ref{eq:trunc}), 
one
obtains the corresponding solid lines in  Fig.~\ref{fig1}. The comparison
shows that the effect of these hole-hole ladders on the calculated energy of
nuclear matter is rather weak in this range of densities. 
The hole-hole ladders yield an effect which is
weakly repulsive. The effect is a little bit larger for the V14 interaction as
compared to the softer CDBONN. All these results indicate that the coupled
cluster approach, restricted to the $S_m$ amplitudes with $m\leq2$ 
yields results very similar to the BHF approximation, employing the
conventional choice for the intermediate particle spectrum. 

However, it is not the issue of the present work to perform nuclear structure
studies within the conventional approach. More sophisticated calculations
including up to three-hole line terms in the Brueckner expansion scheme for
nuclear matter\cite{baldo} or coupled-cluster calculations for finite nuclei
including $S_3$ terms have been performed\cite{zabol,heisenb}. 

The central aim of this work is to account for the isobar excitations
using the coupled-cluster approach restricted to $m$ less or equal 2. 
For that purpose we consider the baryon baryon interaction models
V28 \cite{V28}, Bonn$_{2000}$\cite{mac00} and the Chiral Quark Cluster (CQC) 
model developed in Salamanca\cite{entem},
which all include the scattering to $N\Delta$ and $\Delta\Delta$ states.
While V28 and Bonn$_{2000}$ yield rather accurate fits of the NN
phase shifts in all partial waves, the CQC model leads to such a good fit
only for the channels with isospin T=0, J=1 and T=1, J=0 channels. Therefore we
have replaced the CQC model by the Bonn$_{2000}$ interaction model in all
other channels.

For the case of nuclear matter the equations (\ref{eq:trunc}) has been solved as
an integral equation employing the usual angle-average approximation for the
Pauli operator\cite{haftel}.  The hole-hole ladder term can be introduced as an
additional non-linear term, for which self-consistency is obtained in an
iterative procedure. 

In the case of finite nuclei we have solved the coupled-cluster equation by
considering an expansion of the correlated two-body wave function in a basis
of relative wave function defined in a box of a given radius. This basis
provides an independent control of the maximal distance and momentum relevant
for the correlated waves. The method, restricted to nucleon-nucleon correlations
only has been described in \cite{stauf}. The extension to include isobar
configurations is straight forward\cite{frick1}.

As a result of the calculation we not only obtain the energy of the system
(\ref{eq:ener}) but one can also calculate the one-body density matrix
\begin{equation}
\rho_{\alpha\beta} = \frac{\matrixel{\Psi}{a^\dagger_\alpha
a_\beta}{\Psi}}{\left< \Psi \vert \Psi \right>} \,,\label{eq:onedens}
\end{equation} 
using the techniques described in \cite{emrich}. This one-body density allows
for the evaluation of the radius of finite nuclei and also yields the
probability that a $\Delta$ is excited.

\section{Results and Discussion}  
 
\begin{figure}
\begin{center}
\epsfig{figure=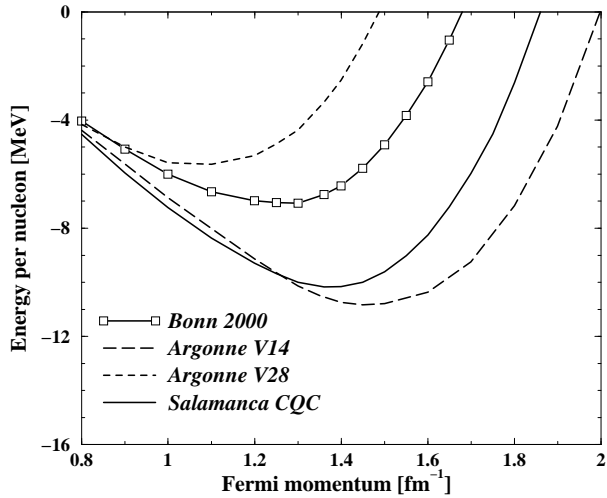,width=8cm}
\end{center}
\caption{Binding energy of nuclear matter as a function of the Fermi momentum. 
The contributions from hole-hole ladders, which are negligibly small 
(see Fig.~1), have been ignored in the results displayed here.
Results are given for the various interaction models with explicit consideration
of isobar excitations. For a comparison the result obtained for the conventional
Argonne V14 interaction is also included.\label{fig2}}
\end{figure}

The energy per nucleon calculated for homogenous nuclear matter at various
densities is displayed in Fig.~\ref{fig2}. The results obtained for the three
different interactions treating isobar excitations explicitely (Argonne V28,
Bonn$_{2000}$ and Salamanca CQC) are compared to those obtained within the
conventional framework using the Argonne V14 interaction model. All results have
been obtained in the coupled-cluster or exponential S approach restricting the
excitation operator to $S_m$ with $m\leq 2$. All calculations including isobar
configurations yield results for the binding energy, which are less attractive
than the result
obtained for the conventional calculation. 

The reason for this loss of binding energy has been presented already long time
ago (see e.g.\cite{anast}) and we just want to repeat it using the language of
perturbation theory. The contribution of isobar configurations to the effective
interaction of 2 nucleons can be written in lowest perturbation theory as
\begin{equation}
\Delta V = V_{N\Delta}^\dagger \frac{Q}{\omega - H_{N\Delta}} V_{N\Delta} + 
V_{\Delta\Delta}^\dagger \frac{1}{\omega - H_{\Delta\Delta}} V_{\Delta\Delta}\,,
\label{deltav}
\end{equation} 
where $V_{N\Delta}$  and $V_{\Delta\Delta}$ represent the transition potentials
for the $NN\to N\Delta$ and $NN \to \Delta\Delta$ transitions, respectively. The
energy of the interacting nucleons is denoted by $\omega$ and $H_{N\Delta}$ (
$H_{\Delta\Delta}$ describe the hamiltonian for the intermediate $N\Delta$
($\Delta\Delta$) states. These contributions are responsible for a sizeable part
of the medium range attraction of the $NN$ interaction. Therefore a realistic
interaction model like the Argonne V14, which does not allow for isobar
configurations contains attractive components which simulate the effects of
isobar excitations in fitting the NN scattering phase shifts. The effects of the
attractive isobar terms displayed in (\ref{deltav}) are reduced in the nuclear
medium, since the interacting nucleons are bound ($\omega$ gets negative) and
because a part of the $N\Delta$ configurations is unavailable due to the Pauli
blocking. This quenching of the attractive isobar terms is observed only, if the
isobar effects are treated explicitely, it is not contained in the conventional
models, that simulate the isobar effects in the effective NN interaction in a
pure phenomenological way.

These arguments explain the loss of attraction due to the explicit inclusion of
isobar excitations. Since the reduction of the isobar terms increases with
density, they can also explain that the repulsion increases with increasing
density, a feature which tends to shift the saturation point to lower densities.

\begin{figure}
\begin{center}
\epsfig{figure=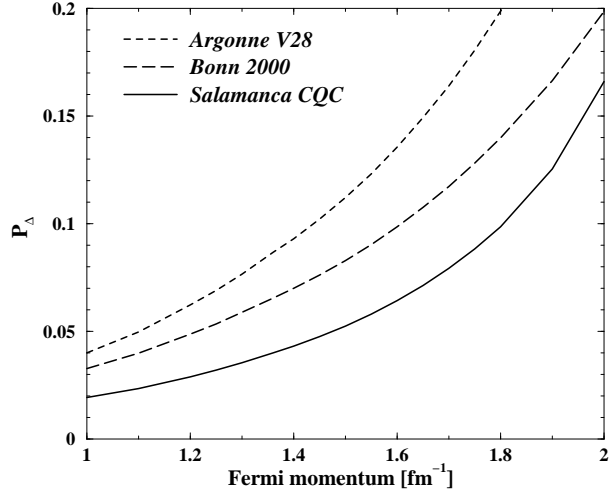,width=8cm}
\end{center}
\caption{$\Delta$ probability per nucleon in nuclear matter as a function of the
Fermi momentum. Results are presented for the various interaction
models discussed in text.\label{fig3}}
\end{figure}

The calculated binding energy is smallest for the Argonne V28 interaction model
and slightly larger for the Bonn$_{2000}$ and the Salamanca CQC model. To some
extent this could be explained by the observation that local interaction models,
like Argonne V28, tend to predict a weaker binding than non-local
interactions, which fit the same NN phase shifts\cite{local}. This feature,
however, may also be interpreted as an indication that the predicted isobar 
effects are larger for Argonne V28 than for the other two models under
consideration. This interpretation is supported by the calculated probabilities
of isobar excitations in nuclear matter, displayed in Fig.~\ref{fig3}. 

The prediction for the isobar probabilities derived from the various interaction
models differ in a very significant way. At the empirical saturation density,
which corresponds to a Fermi momentum $k_F$ = 1.36 fm$^{-1}$, the difference is
larger than a factor two (see also Table~\ref{tab1}).  Despite these differences
in the prediction of the total $\Delta$ probability, there are some common
features in the predictions of these models with quite different origin. If one
tries to analyze, which partial waves provide the most important contributions,
one observes that all interaction models predict a larger contribution from the
excitation of $\Delta\Delta$ configurations, which can occur in interacting
pairs of baryons with $T=0$ and $T=1$, than from the excitation of $N\Delta$
excitations, which occur in $T=1$ partial waves only.

All interaction models predict large contributions from those partial waves, in
which the interacting nucleons are in a state with relative angular momentum
$l=0$. However, for a complete calculation one cannot ignore the contributions
from the higher partial waves (see also Table~\ref{tab1}).

The features we have discussed so far for the case of infinite nuclear
matter are also observed in the results obtained for the finite nucleus
$^{16}$O, which are displayed in Table~\ref{tab2}. The calculated binding
energy per nucleon is smallest for the Argonne V28, about 0.7 MeV  and
2.4 MeV per nucleon larger for the Bonn$_{2000}$ and Salamanca model,
respectively. The increase in the calculated binding energy is correlated with
a smaller prediction for the $\Delta$ probability $P_\Delta$. Also for $^{16}$O
we observe a difference by almost a factor of 2 between $P_\Delta$ derived from
V28 and Salamanca CQC model. The probabilities $P_\Delta$ calculated for
$^{16}$O are similar in magnitude as those evaluated for nuclear matter at
small densities ($k_F$ around 1 fm$^{-1}$). Therefore a nuclear matter
calculation of isobar effects seems to provide a reasonable first estimate for
the case of  finite nuclei, if one uses a local density approximation. For a
more detailed information, like the relative importance of different partial
waves, an explicit calculation of the finite systems is required.

\begin{figure}
\begin{center}
\epsfig{figure=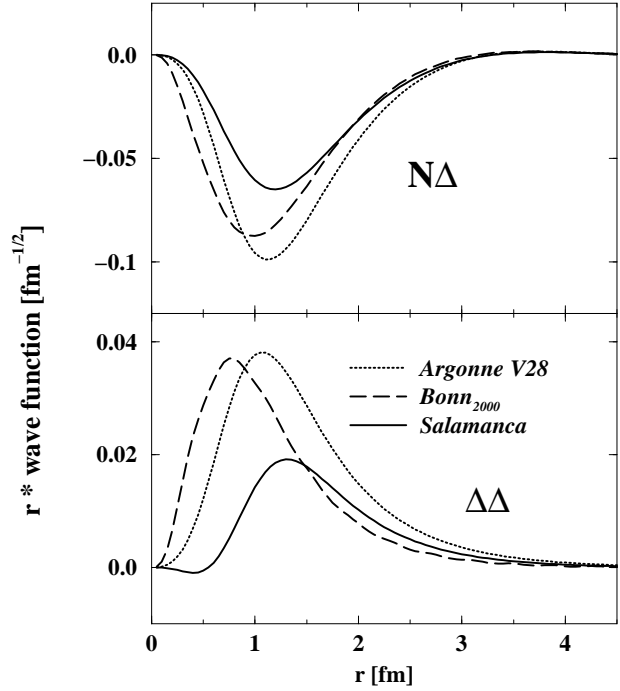,width=8cm}
\end{center}
\caption{$N\Delta$ (upper part) and $\Delta\Delta$ correlation function
originating from two nucleons in the $0s_{1/2}$ shell of $^{16}O$ as a function
of the relative distance $r$.\label{fig4}}
\end{figure}

The differences in the calculated energies and $\Delta$ probabilities obtained
in these different interaction models originate of course mainly from the
different transition potentials $V_{N\Delta}$ and $V_{\Delta\Delta}$ (see also
eq.(\ref{deltav})). Some of these differences, like the treatment of the $\pi$
exchange contribution or the short-range behaviour in the local V28 approximation
as compared to the
non-local calculation in the other models  have been discussed already in
\cite{frick1}. These differences are also the main origin for the different
predictions obtained in the present calculation. As a
typical example we show the amplitudes 
\begin{equation}
\left< N\Delta\, ^5D_0\vert S_2 \vert 0s_{1/2}0s_{1/2} \right>_{J=0,T=1}\,,
\label{examp:s2}
\end{equation}
calculated for $^{16}$O as a function of the relative distance $r$ of the
$N\Delta$ pair in the upper part of Fig.~\ref{fig4}. The lower part contains the
corresponding correlation function for the $\Delta\Delta$ configuration. 
Inspecting the differences obtained from the three interaction models, one can
see that the V28 interaction leads to a larger amplitude at larger distances
$r$ than the other two. This can be related to the fact that V28 uses a local 
$\pi$ exchange contribution in the transition potential, which does not account
for retardation effects which are due to the $N\Delta$ mass difference. The
different behaviour in the correlation functions $S_2$ at small distances $r$
must be related to the different kinds of models. While the short-range
behaviour in the V28 and Bonn$_{2000}$ model are controlled by local and
non-local form-factors, respectively, it is the coupled channel calculation
within the chiral quark model, which provides the short-range behaviour in the
Salamanca model.

\section{Conclusions}

In the present investigation we try to compare the predictions for the bulk
properties of nuclei derived from three different baryon-baryon interaction
models, which  account for isobar configurations explicitely. The main
differences can be related to the models for transition potentials describing
$NN \to N\Delta$ and $NN \to \Delta\Delta$ transitions. The quark model of the
Salamanca CQC approach predicts weaker transition amplitudes at short range
than the more phenomenological cut-offs in the Bonn$_{2000}$ and Argonne V28
interaction. The long-range components of these amplitudes are dominated by the
$\pi$-exchange, which is weaker in the non-local models (Salamanca and
Bonn$_{2000}$) than in the local interaction model (V28). 

These differences in the interactions are responsible for the differences in
the predicted  $\Delta$ probabilities in $^{16}$O, which vary between 1.98
percent derived for the Salamanca model and 3.87 percent for Argonne V28. The
results are in fair agreement with the estimates derived from
experiment\cite{exp98}. A
large probability for isobar excitations is related to weak binding energy.
Similar results are obtained for infinite nuclear matter. The isobar
effects discussed here would correspond to the inclusion of a repulsive
three-nucleon force in conventional nuclear structure calculations.   

Comparing the calculated binding energies with empirical values, one must keep
in mind that the coupled-cluster approximation (including terms up to $S_2$),
essentially corresponds to the Brueckner-Hartree-Fock approximation with a
conventional choice of the intermediate particle spectrum. More binding
energy is obtained from including three-body terms or using the so-called
continuous choice spectrum. This is known from corresponding calculation using
nucleon degrees of freedom only\cite{baldo}. The explicit treatment of isobar
configuration, however, would give rise to additional three-body terms, which
are not yet taken into account. The effects become very large at higher
densities if one
considers old local models for the isobar excitations\cite{pico3,dring}. The
isobar effects may provide reasonable corrections if the modern interaction
models are considered. 

We like to acknowledge financial support from the {\it Europ\"aische
Graduiertenkolleg T\"ubingen - Basel} (DFG - SNF), the DGICYT (Spain)
under contract PB98-1247, the US National Science Foundation under
Grant No.~PHY-0099444, and the Ram\`on Areces Foundation (Spain).

\begin{table}
\begin{center}
\begin{tabular}{c|rrr}
&\multicolumn{1}{c}{$\mbox{V}_{28}$} &
\multicolumn{1}{c}{$\mbox{Bonn}_{2000}$} & \multicolumn{1}{c}{CQC}
 \\ \hline
$P_{\Delta}$ (total) & 8.67 & 6.55 & 4.00 \\
$P\, (N\Delta)$ & 4.03 & 2.79 & 1.87 \\
$P\, (\Delta\Delta )$ & 4.64 & 3.76 & 2.13 \\
$P_\Delta (^1S_0)$ & 2.77 & 2.43 & 1.24 \\
$P_\Delta (^3S_1)$& 1.80 & 2.00 & 0.64 \\
\end{tabular}
\caption{\label{tab1} $\Delta$ probability per nucleon in nuclear matter at 
saturation density
derived from various interaction models. The total probability originates from
the excitation of $N\Delta$ ($P\, (N\Delta)$) and $\Delta\Delta$ configurations.
Also shown are predictions if only the coupling to $\Delta$ configuration from
$NN$ channels $^1S_0$ and $^3S_1$ are considered. All entries are in percent.}
\end{center}
\end{table} 

\begin{table}
\begin{center}
\begin{tabular}{c|rrr}
&\multicolumn{1}{c}{$\mbox{V}_{28}$} &
\multicolumn{1}{c}{$\mbox{Bonn}_{2000}$} & \multicolumn{1}{c}{CQC}
 \\ \hline
E/A [MeV] & -2.73 & -3.49 & -5.15 \\
r [fm] & 2.81 & 2.65 & 2.53 \\
\hline
$\varepsilon_{s1/2}$ [MeV] & -34.65 & -37.88 & -44.25 \\
$\varepsilon_{p3/2}$ [MeV] & -16.49 & -18.54 & -22.03 \\
$\varepsilon_{p1/2}$ [MeV] & -13.56 & -14.82 & -17.67 \\
\hline
$P_\Delta$ [\%] & 3.87 & 3.71 & 1.97 \\
$P_\Delta$ (N$\Delta$) [\%] & 1.75 & 1.43 & 0.88 \\
$P_\Delta$ ($\Delta\Delta$) [\%] & 2.12 & 2.28 & 1.08 \\
\end{tabular}
\caption{Energy per nucleon (E/A), radius (r), single-particle energies for the
nucleons and $\Delta$ probability per nucleon for $^{16}$O. The total
probability originates from the excitation of $N\Delta$ ($P\, (N\Delta)$) and
$\Delta\Delta$ configurations. Results are presented for the various interaction
models discussed in text.\label{tab2}}
\end{center}
\end{table}

\end{document}